\documentclass{article}
\usepackage[cp1251]{inputenc}
\usepackage{amsmath}
\usepackage[russian]{babel}

\usepackage{russ}
\usepackage{amsmath}
\usepackage{amsfonts}
\usepackage{amssymb}
\usepackage{graphicx}
\usepackage{amssymb,amscd}
\usepackage[cp1251]{inputenc}
\usepackage[T2A]{fontenc}
\usepackage[russian]{babel}

\usepackage{yfonts}
\usepackage{amsmath}
\usepackage{amsfonts}
\usepackage{amssymb}
\usepackage{graphics}
\usepackage{amssymb,amsmath,mathrsfs,amsthm,amscd}

\newtheorem{thm}{Теорема}

\newtheorem{zam}{Замечание}
\newtheorem{lemm}{Лемма}
\newtheorem{definition}{Определение}
\begin{document}

\title
{Спектральные серии оператора Шредингера с дельта-потенциалом на трехмерном сферически симметричном многообразии\footnote{Работа выполнена на при поддержке гранта правительства РФ для господдержки научных исследований, проводимых под руководством ведущих ученых, в ФГБОУ ВПО “Московский Государственный Университет имени М.В. Ломоносова” по договору N 11G.34.31.0054, а также грантов РФФИ 11-01-00937-а, 13-01-00664, программы поддержки ведущих научных школ (грант НШ-3224.2010.1) и
гранта поддержки молодых ученых “Мой первый грант"12-01-31235.}.}
\author{Тудор С. Ратью, А.А. Сулейманова, А.И. Шафаревич}
\date{}
\maketitle

\begin{abstract}
The spectral series of the Schrцdinger operator with a delta-potential on a three-dimensional compact spherically symmetric manifold in the semiclassical limit as $h\to0$ are described.
\end{abstract}

\section{Введение}

Исследованию операторов Шрёдингера с дельта-потенциалами (точечными потенциалами, потенциалами нулевого радиуса) посвящено много физических и математических работ. Модель точечных потенциалов может использоваться для описания короткодействующих примесей, дефектов и подобных явлений в различных системах. Одной из первых работ, в которых потенциалы нулевого радиуса применялись для исследования зонного спектра периодических систем, является статья \cite{Kronig}, где рассматривалась модель нерелятивистского электрона, движущегося в жесткой кристаллической решетке. С тех пор модель приобрела значительную популярность, особенно в атомной и ядерной физике(см., например, \cite {bethe}, \cite {goldb}, \cite {zeld}, \cite {fas1}, \cite {fas2}, \cite {fas3}, \cite {krevch}, \cite {peng}, \cite {krev}).

Строгое математическое обоснование метода дельта-потенциалов было дано в работе \cite{Berezin}, где было предложено использовать формулу М.Г. Крейна для описания резольвент операторов с точечными возмущениями. Обширная библиография работ, посвящённых применениям метода точечных потенциалов содержится в монографиях \cite {paper1}, \cite {paper2}. В работах \cite {Bruening}, \cite {paper4}, \cite {paper5}, \cite {paper15} на основе теории расширений изучались спектральные свойства операторов с дельта-потенциалами и близких к ним операторов на сингулярных пространствах.

В данной работе описаны спектральные серии оператора Шредингера с дельта-потенциалом вида $ H=-\frac{h^2}2 \Delta+ \alpha\delta_{x_0}(x), \alpha\in\mathbb {R} $ в квазиклассическом пределе $h \to 0$ на трехмерной компактной поверхности, обладающей сферической симметрией. Для широкого класса уравнений с гладкими коэффициентами квазиклассическая теория развита в работах В.П. Маслова (см.,например, \cite {maslov2}, \cite {Maslov}); в частности, из них вытекает следующий результат. Пусть $N$ --- риманово многообразие и $V: N\to\mathbb{R}$ --- гладкая функция (потенциал). Если гамильтонова система в $T^*N$, задаваемая гамильтонианом $\frac12|p|^2+V$ вполне интегрируема, то соответствующие лиувиллевы торы $\Lambda$ определяют квазиклассические спектральные серии оператора $H=-\frac{h^2}2\Delta +V(x)$ (здесь $x\in N$, $(x,p)$ -- стандартные координаты на $T^*N$). Именно: асимптотика при $h\to 0$ собственных чисел оператора $H$ вычисляется из условий квантования Бора --- Зоммерфельда --- Маслова
\begin{equation}\label{bz}
\frac1{2\pi h}\int_\gamma(p,dx)+\frac14\mu(\gamma)=m\in\mathbb{Z},
\end{equation}
где $\gamma$ --- произвольный цикл на $\Lambda$, $\mu$ -- индекс Маслова, $m=O(1/h)$. Формальная асимптотика собственных функций (квазимоды) имеет вид $\psi=K_{\Lambda}(1)$, где $K_\Lambda$ --- канонический оператор Маслова на торе $\Lambda$, удовлетворяющем условию квантования.

К операторам с дельта-потенциалами конструкция канонического оператора, вообще говоря, неприменима; геометрия соответствующей классической задачи к настоящему времени остается мало исследованной. Ниже описаны инвариантные лагранжевы многообразия, соответствующие спектральным сериям указанного оператора с дельта-потенциалом и получены условия квантования, определяющие асимптотику собственных значений. Эти условия, вообще говоря, нестандартны; при больших или малых значениях коэффициента $\alpha$ они переходят в равенства вида \eqref{bz}, но с разными значениями индекса Маслова $\mu$ --- возможно, это указывает на наличие более сложных геометрических объектов, связанных с квазиклассической теорией операторов с сингулярными коэффициентами. Работа представляет собой продолжение работ \cite{Filatova}, \cite{Ratiu}; в них аналогичная задача изучалась для стандартной сферы (в этом случае спектр вычисляется точно) и для двумерной поверхности вращения.

\section{Постановка задачи.}
\subsection{Спектральная задача.}

Будем рассматривать спектральную задачу
\begin{equation} \label{main}
\left(- \frac{h^2}2 \Delta+\alpha\delta_{x_0}(x))\right) \Psi = E\Psi ,\quad x\in N,\quad \alpha\in\mathbb {R},
\end{equation}
где $\delta_{x_0}(x)$ - дельта-функция Дирака, сосредоточенная в точке $x_0$, в квазиклассическом пределе $h \to 0$ на трехмерном многообразии в $\mathbb{R}^4$
$$
N=(f(z)\cos\theta \cos\varphi, f(z)\cos\theta \sin \varphi, f(z)\sin\theta, z),
$$
где $z\in[z_0,z_1]$, $0\le\varphi\le2\pi$, $-\frac{\pi}2\le\theta\le\frac{\pi}2$. Относительно функции $f(z)$ будем предполагать следующее.

\begin{enumerate}
\item $f(z_0)=f(z_1)=0$, $f(z)>0$ при $z\in (z_0,z_1)$;
\item $f(z)=\sqrt{(z_1-z)(z-z_0)}\omega(z)$, где $\omega(z)$ --- многочлен.
\end{enumerate}

При этих условиях поверхность $N$ --- аналитическое многообразие, диффеоморфное трехмерной сфере; точки $x_0,x_1$, соответствующие значениям параметра $z=z_0,z_1$ --- полюса этой поверхности (в одном из них сосредоточена дельта-функция).

\begin{zam}
Условие 2 можно ослабить --- по-видимому, достаточно требовать аналитичность $f$ в некоторой окрестности отрезка $[z_0,z_1]$, за исключением точек $z_0,z_1$, в которых $f$ имеет корневую особенность.
\end{zam}

Ниже приведено формальное определение оператора с $\delta$-потенциалом на поверхности $N$.

\subsection{Формальное определение оператора $H$.}

Оператор

\begin{equation} \label{H}
H=-\frac{h^2}2 \Delta + \alpha\delta_{x_0}(x),\quad x\in N,\quad\alpha\in\mathbb{R}
\end{equation}
в пространстве $L_2(N)$ определяется с помощью конструкции самосопряженных расширений (см.\cite{Berezin}). Именно: $H$ строится таким образом, чтобы выполнялись следующие требования.

\begin{enumerate}
\item Оператор $H$  самосопряжен.
\item На функциях, обращающихся в ноль в точке $x_0$, $H$ совпадает с оператором $H_0=-\frac{h^2}{2}\Delta$, где $\Delta$ --- оператор Лапласа --- Бельтрами.
\end{enumerate}

Точнее: рассмотрим самосопряженный оператор $H_0$ с областью определения $D(H_0)=W^2_2(N)$, где $W^2_2(N)$ --- второе пространство Соболева. Ограничим действие оператора $H_0$ на функции $\psi(x)$, такие что $\psi(x_0) = 0$; получим симметрический оператор  $H_0|_{\psi(x_0)=0}$.

\begin{definition}
Оператором $H=-\frac{h^2}2 \Delta + \alpha\delta_{x_0}(x)$ называется самосопряженное расширение оператора $H_0|_{\psi(x_0)=0}$.
\end{definition}

\begin{zam}
Все такие расширения параметризуются одним вещественным параметром $\alpha$, который естественно трактовать как коэффициент при $\delta$-потенциале в (\ref{H}) (в частности, при $\alpha=0$ получаем $H=H_0$).
\end{zam}

Каждое расширение задается граничными условием в точке $x_0$; точнее, область определения оператора $H$ состоит из функций следующего вида
$$
\psi=\psi_0+c_1G(x,x_0;i)+с_2G(x,x_0,-i),
$$
где $\psi_0\in W_2^2(N)$, $\psi_0(x_0)=0$, $G(x,y,\lambda)$ --- функция Грина оператора $\Delta$, т.е. интегральное ядро резольвенты:
$$
(\Delta-\lambda)^{-1}f = \int_M G(x,y;z)f(y)\Omega
$$
($\Omega$ --- форма объема на $N$).

В точке $x_0$ функции указанного вида имеют особенность; именно, справедливо следующее разложение
\begin{equation} \label{razlozh}
\psi(x)=-\frac{a}{4\pi}{d(x,x_0)}^{-1}+b+o(1),
\end{equation}
где $a, b \in \mathbb {C}, d(x,x_0)$ --- геодезическое расстояние между $x$ и $x_0$ на $N$.
Область определения расширения $H$, соответствующего параметру $\alpha$,  состоит из функций, удовлетворяющих граничному условию
\begin{equation} \label{uslovie}
a=\frac{2\alpha}{h^2}b.
\end{equation}

\section{Формулировка результата.}

\subsection{Описание лагранжева многообразия.}

Квазиклассическая асимптотика собственных чисел оператора $H$ вычисляется из условия квантования на лагранжевом многообразии, которое мы сейчас опишем. Пусть $(x,p) \in T^*N$, где $T^*N$ --- кокасательное расслоение к $N$, $x\in N$, $p$ --- кокасательный вектор к $N$ (импульс).
Рассмотрим гамильтонову систему (геодезический поток)
\begin{equation} \label{ham}
\left\lbrace
 \begin{aligned}
 &\dot x=\frac {\partial \mathcal{H}} {\partial p};\\
 &\dot p=-\frac {\partial \mathcal{H}} {\partial x},
 \end{aligned}
 \right.
\end{equation}
где $\mathcal{H}=\frac{|p|^2}2$, и ее траектории
$$
\left\lbrace
 \begin{aligned}
 &x=X(\omega,t);\\
 &p=P(\omega,t), \,\,\,\, \omega \in S^2_{\sqrt{2E}}, \,\, t \in \mathbb{R},
 \end{aligned}
 \right.
$$
заданные начальными условиями
\begin{equation} \label{usl}
\left\lbrace
 \begin{aligned}
 &x(0)=x_0;\\
 &p(0)=\omega, \,\,\,\, \omega \in S^2_{\sqrt{2E}}, \,\, |\omega|=\sqrt{2E}.
 \end{aligned}
 \right.
\end{equation}
Здесь $S^2_{\sqrt{2E}}$ ---- двумерная сфера радиуса $\sqrt{2E}$ в кокасательном пространстве в точке $x_0$. Таким образом,
из точки $x_0$ (в которой сосредоточен дельта-потенциал) по поверхности $N$ с импульсом, "бегающим" по сфере $S^2_{\sqrt{2E}}$ (т.е. $p \in \Lambda_0$, где $\Lambda_0=\{x=x_0, |p|=\sqrt{2E}\}$), выпускаются траектории гамильтоновой системы. Эти траектории лежат на многообразии $\Lambda=\bigcup_t g_t\Lambda_0$ ($g_t$ --- гамильтонов фазовый поток), диффеоморфном $S^2 \times S^1$ (см. Рис. 1); оно описывает классические движения, соответствующие данной квантовой задаче. Проекции траекторий на $N$ --- геодезические.

\begin{figure}[h]
\includegraphics[scale=0.55]{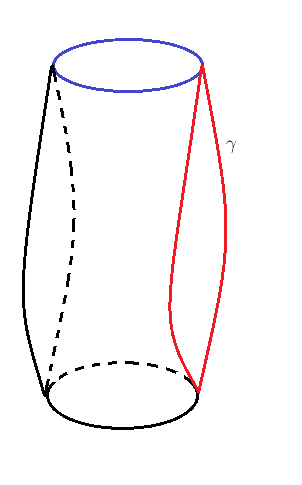}

Рис. 1: Лагранжево многообразие, соответствующее задаче (\ref{main}).
\end{figure}

\begin{figure}[h]
\includegraphics[scale=0.65]{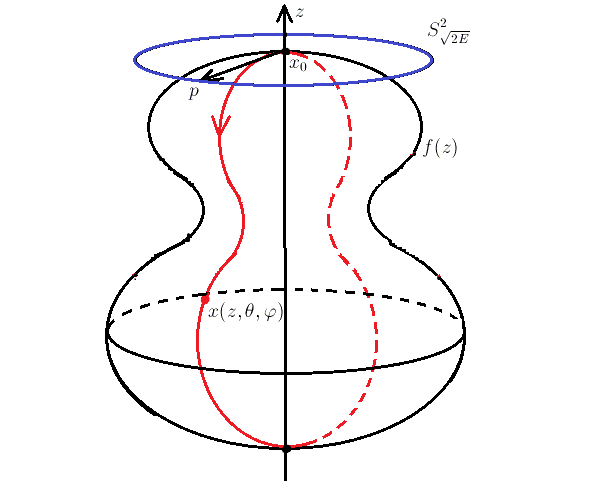}

Рис.2: Классические траектории на поверхности $N$, соответствующие квантовой задаче (\ref{main}).
\end{figure}

На многообразии $\Lambda$ имеется один базисный цикл $\gamma$, интегралы по нему и дают вклад в спектр. В качестве этого цикла можно взять замкнутую траекторию гамильтоновой системы (\ref{ham}) с начальными условиями (\ref{usl}) (см. Рис. 2).

Проекция $\Lambda$ на $x$-пространство устроено следующим образом: в каждую точку $N$, кроме $x_0$ и $x_1$, проецируется две точки многообразия $\Lambda$ вида $(x,p)$ и $(x,-p)$. В каждую из точек $x_0$ и $x_1$ проецируется двумерная сфера.

\subsection{Формулировка результата: условие квантования.}

Условие квантования на многообразии $\Lambda$ по циклу $\gamma$ и есть искомое уравнение на спектр задачи; точнее, справедливо следующее утверждение.

\begin{thm}
Пусть $Ch^{-\epsilon}<\frac{\alpha}{h^3}<Ch^\epsilon$ для некоторого достаточно малого $\epsilon$. Пусть существует число $E=O(1)$, удовлетворяющее условию квантования
$$
\tan(\frac1 {2h} \oint\limits_{\gamma}(p,dx)+O(h)) = \frac{2h^3}{\sqrt{2E}\alpha},
$$
где $\gamma$ --- указанный выше цикл (замкнутая траектория) на лагранжевом многообразии $\Lambda$. Тогда существует собственное значение $E_0$ оператора $H$, такое что $|E-E_0|=o(h)$, при $h\to 0$.
\end{thm}

\begin{zam}
Явный аналитический вид условия квантования таков

$$
\tan(\frac1h \int_{z_0}^{z_1}\sqrt{2E(f'^2+1)}dz+O(h)) = \frac{2h^3}{\sqrt{2E}\alpha}
$$
\end{zam}

\begin{zam}
Асимптотика собственной функции вне сколь угодно малой не зависящей от $h$ окрестности точки $x_0$ имеет вид $K_{\tilde\Lambda}(1)$, где $K$ --- канонический оператор Маслова, а $\tilde\Lambda$ --- некомпактное лагранжево многообразие, полученное из $\Lambda$ выбрасыванием двумерной сферы, проецирующейся в точку $x_0$ (это многообразие, очевидно, гомеоморфно цилиндру $S^2\times\mathbb{R}$.
\end{zam}

\subsection{Скачок индекса Маслова.}
Рассмотрим предельные случаи условия квантования, описанного в теореме. Пусть $\frac{\alpha}{h^3}\to 0$, тогда условия квантования принимают стандартный вид
$$
\frac1{2\pi h}\oint\limits_{\gamma}(p,dx)=k+\frac12
$$
(отметим, что индекс Маслова цикла $\gamma$ равен двум).
Пусть теперь $\frac{\alpha}{h^3}\to \infty$; тогда имеем
$$
\frac1{2\pi h}\oint\limits_{\gamma}(p,dx)=k.
$$
Это равенство также имеет вид условия Бора --- Зоммерфельда --- Маслова; однако в этом случае ``индекс Маслова'' цикла $\gamma$ оказывается равным нулю. Таким образом, при переходе через критическое значение $\alpha=O(h^3)$ происходит скачок целочисленного инварианта, совпадающего в случае гладкого потенциала с индексом Маслова: наличие дельта-функции приводит к его изменению на 2. Возможно, это указывает на существование некоторой топологической конструкции (пока нам неясной), обобщающей  канонический оператор Маслова на случай сингулярных коэффициентов.

\begin{zam}
Теорема остается справедливой и без сформулированного ограничения на $\alpha$. Однако вне указанного диапазона условие квантования всегда имеет вид правила Бора --- Зоммерфельда --- Маслова; кроме того, при доказательстве теоремы может, вообще говоря, потребоваться построение нескольких поправок к асимптотике (соответствующая процедура аналогична описанной ниже), причем количество поправок зависит от порядка отношения $\alpha/h^3$.

\end{zam}

Остальная часть работы посвящена доказательству теоремы 1.

\section{Доказательство теоремы 1.}

\subsection{Разделение переменных.}
Асимптотическое решение задачи
\begin{equation}\label{main2}
H\psi(x)=E\psi(x)+o(h)
\end{equation}
строим в виде
\begin{equation}\label{solution}
\psi(x)=e_1\psi_1(x)+e_2\psi_2(x),
\end{equation}
где $e_1, e_2$ --- разбиение единицы $e_1+e_2=1$, причем носитель функции $\psi_1(x)$ лежит вблизи $x_0$, носитель функции $\psi_2(x)$ вне $x_0$. В точках пересечения носителей функции должны совпадать $\mathrm{mod} o(h)$.

Метрика на поверхности вращения $N$ имеет вид
$$
g_{ij}=
\begin{pmatrix}
f'^2(z)+1 &0&0\\
0&f^2(z)&0\\
0&0& \cos^2\!{\theta}f^2(z)
\end{pmatrix}
, \sqrt{g}=f^2(z)\sqrt{f'^2(z)+1}\,\cos{\theta}.
$$
Оператор Лапласа в координатах записывается следующим образом
$$
\Delta = \frac1 {f^2(z)\sqrt{f'^2(z)+1}} \frac \partial {\partial z} \frac{f^2(z)}{\sqrt{f'^2(z)+1}} \frac \partial {\partial z} + \frac 1 {f^2(z)} \Delta_0,
$$
где $\Delta_0$ --- оператор Лапласа на двумерной сфере:
$$
\Delta_0= \frac 1 {\cos{\theta}} \left( \frac \partial{\partial\theta} \cos{\theta} \frac \partial {\partial\theta} + \frac \partial {\partial\varphi} \frac 1 {\cos{\varphi}} \frac \partial {\partial\varphi} \right).
$$

Будем искать решение (\ref{solution}) в виде $\psi = u(z)Y$, где $-\Delta_0 Y=m(m+1)Y$. После подстановки уравнение (\ref{main2}) будет выглядеть так
\begin{equation}\label{equ}
u''(z) + \left( \frac{2f'(z)}{f(z)} - \frac{f'(z)f''(z)}{f'^2(z)+1} \right) u'(z) + (f'^2(z)+1) \left( \frac{2E}{h^2} - \frac {m(m+1)}{f^2(z)} \right) u(z) = 0.
\end{equation}

\subsection{Структура решений в окрестности особых точек.}

В уравнении \eqref{equ} две регулярных особых точки: $z_0$ и $z_1$. Выясним локальную структуру решений в окрестности этих точек; пусть, для определенности, $z\to z_0$.
Подставим разложение $f(z)=\sqrt{z-z_0}\,\sum^\infty_{j=0} a_j(z-z_0)^j$ в уравнение (\ref{equ}) и учтем только члены порядка $(z-z_0)^{-1}$. Получим
\begin{equation}\label{main3}
u''(z)- \frac2 {z_0-z}u'(z) + \frac{a_0^2}{4(z_0-z)}\left( \frac{2E}{h^2} - \frac{m(m+1)}{a_0^2(z_0-z)} \right)u(z)=0.
\end{equation}

Для вычисления характеристических показателей (см., например, \cite{Fedoruk}), подставим в уравнение (\ref{main3})
$$
u(z)=(z-z_0)^\rho r(z),
$$
где $r(z)=\sum^{\infty}_{k=0}{r_k(z-z_0)^k}$. Приравняем к нулю коэфициент при $(z-z_0)^{\rho-2}$, получим
$$
\rho(\rho-1)+2\rho- \frac{m(m+1)} 4 =0,
$$
то есть $\rho =\frac{-1\pm\sqrt{1+m(m+1)}} 2$, и, пользуясь \cite{Fedoruk}, получаем, что \eqref{equ} имеет два линейно независимых решения следующего вида
$$
\left\lbrace
 \begin{aligned}
 R_1(z)=(z-z_0)^{\rho_1} \varphi_1(z);\\
 R_2(z)=(z-z_0)^{\rho_2} \varphi_2(z),
 \end{aligned}
 \right.
$$
где $\varphi_1(z), \varphi_2(z)$ голоморфны в окрестности $z_0$.

Нам нужно найти решение \eqref{equ}, аналитическое в окрестности точки $z_1$ и имеющее особенность вида $(z-z_0)^{-1}$ в окрестности точки $z_0$. Ясно, что такие решения существуют лишь при $m=0$; этим случаем мы в дальнейшем и ограничимся. В этом случае имеем

$$
\left\lbrace
 \begin{aligned}
 R_1(z)=\varphi_1(z);\\
 R_2(z)=(z-z_0)^{-1} \varphi_2(z).
 \end{aligned}
 \right.
$$
В окрестности точки $z_1$ решения устроены аналогично.

\begin{zam}
Если $m\neq 0$, уравнение \eqref{equ} допускает решения, обращающиеся в нуль в точке $z_0$. Такое решение определяет собственную функцию оператора Лапласа --- Бельтрами (без дельта-потенциала); асимптотика в этом случае строится по стандартной схеме \cite{Maslov}
\end{zam}

\subsection{Эталонное уравнение в окрестности $z_0$.}
Согласно формуле для длины кривой на поверхности, заданной параметрически, $d(z,z_0)=\int^{z_0}_{z} \sqrt{f'^2(z)+1}\,\,dz , z \in (z_1,z_0)$, причем
$$
d(z,z_0)\sim(z_0-z)\sqrt{f'^2(z)+1}
$$
вблизи $z_0$.

Исследуем фундаментальную систему решений уравнения (\ref{main3}), учитывая, что $m=0$:
\begin{equation}\label{chast}
P''(z)- \frac2 {z_0-z}P'(z) + \frac{a_0^2}{4(z_0-z)} \frac{2E}{h^2} P(z)=0.
\end{equation}
Сделаем замену $t=\frac{a_0\sqrt{2E}} h \sqrt{z_0-z}$, тогда уравнение (\ref{chast}) принимает такой вид
\begin{equation}\label{Bessel}
P_{tt}(t)+ \frac 3 t P_t(t)+P(t)=0.
\end{equation}
 Это уравнение --- уравнение Бесселя (см., например, \cite{Beitmen}), его решение вблизи точки $z_0$ представляет собой линейную комбинацию функции Бесселя и функции Неймана первого порядка, умноженных на степень аргумента, а именно
\begin{equation}\label{solution1}
P(t)= At^{-1}N_1(t)+Bt^{-1}C_1(t)
\end{equation}

Эти функции мы будем использовать при построении асимптотики решения уравнения \eqref{equ} в окрестности точки $z_0$.

\subsection{Асимптотика решения в окрестности точки $z_0$. Переменная Лангера.}

Теперь рассмотрим общий вид уравнения (\ref{equ})
\begin{equation}\label{equ2}
u''(z)+f'(z) \left( \frac{2f'(z)}{f(z)} - \frac{f'(z)f''(z)}{f'^2(z)+1} \right) u'(z) + (f'^2(z)+1) \frac{2E}{h^2} u(z) = 0.
\end{equation}
Это --- уравнение с двумя регулярными особыми точками $z_0$ и $z_1$. Для удобства обозначим
\begin{equation}\label{p}
p(z)=(z-z_0) \left( \frac{2f'(z)}{f(z)} - \frac{f'(z)f''(z)}{f'^2(z)+1} \right),
\end{equation}
\begin{equation}\label{q}
q(z)=(z-z_0)(f'^2(z)+1),
\end{equation}
тогда уравнение (\ref{equ2}) примет вид
\begin{equation}\label{pq}
(z-z_0)u''(z) + p(z)u'(z) + q(z) \frac{2E}{h^2} u(z)=0,
\end{equation}
где $p(z), q(z)$ аналитические функции при $z\to z_0$.

Пользуясь методом Лангера-Вазова, асимптотическое решение уравнения (\ref{pq}) будем искать в виде
\begin{equation}\label{langer}
u(z)=F(\tau)+\dots,
\end{equation}
где $\tau=\frac{S(z)}{h^2}$ --- переменная Лангера, $S(z)$ --- неизвестная функция, аналитическая в окрестности точки $z_0$. Функции $F$, $S$ выберем так, чтобы $u$ имела особенность нужного вида в точке $z_0$. Подставим (\ref{langer}) в (\ref{pq}), получим
\begin{equation}\label{f0}
\tau F_{\tau\tau} + \frac{p(z)S(z)}{S'(z)(z-z_0)}F_{\tau} + \frac{2ES(z)q(z)}{S'^2(z)(z-z_0)}F+\frac{S''S}{(S')^2}F_\tau+\dots = 0
\end{equation}
Поскольку $S=h^2\tau$, последнее слагаемое мало (оценка получена ниже); отбрасывая его и обозначая
\begin{equation}\label{n}
n(z)=\frac{S(z)}{S'(z)} \left( \frac{2f'(z)}{f(z)} - \frac{f'(z)f''(z)}{f'^2(z)+1} \right),
\end{equation}
\begin{equation}\label{k}
k(z)=\frac{2ES(z)}{S'^2(z)}(f'^2(z)+1),
\end{equation}
приведем уравнение (\ref{f0}) к виду
\begin{equation}\label{nk}
\tau F_{\tau\tau} + n(z)F_{\tau} + k(z)F = 0.
\end{equation}
 Произвольное решение уравнения (\ref{f0}) --- это линейная комбинация следующих функций
\begin{equation}\label{N1}
F_{1}=\tau^{\frac{1-n(z)} 2} N_{n(z)-1}(2\sqrt{k(z)\tau}),
\end{equation}
\begin{equation}\label{C1}
F_{2}=\tau^{\frac{1-n(z)} 2} C_{n(z)-1}(2\sqrt{k(z)\tau}),
\end{equation}
где $N_{\nu}(t)$ -- функция Неймана, $C_{\nu}(t)$ - функция Бесселя порядка $\nu$ (см., например,\cite{Beitmen}).

Для того, чтобы функция $F$ зависела только от $\tau$, величины $k(z)$ и $n(z)$ должны быть константами. Можно считать, что $k(z)=1$; это равенство приводит к уравнению Гамильтона --- Якоби
$$
\frac{2ES(z)}{S'^2(z)}(f'^2(z)+1)=1.
$$
Отсюда находим $S(z)$
$$
S(z)=\left( \frac 1 2 \int^z _{z_0} \sqrt{2E(f'^2(z)+1)} dz \right)^2;
$$
отметим, что эта функция голоморфна в окрестности точки $z_0$ и $S(z_0)=0$, $S'(z_0)\neq 0$.
Функция $n(z)$, вообще говоря, не константа; однако нетрудно понять, что замена в уравнении \eqref{nk} $n(z)$ на $n(z_0)$ приводит к добавлению малого слагаемого. Действительно,
$$
(n(z)-n(z_0))F_\tau=\frac{n(z)-n(z_0)}{S(z)}S(z)F_\tau=h^2\frac{n(z)-n(z_0)}{S(z)}\tau F_\tau
$$
(аккуратная оценка приведена ниже).

Прямое вычисление показывает, что $n(z_0)=2$; таким образом, функцию $F$ представляем в виде
\begin{equation}\label{psi1}
F(\tau)=A_1\tau^{-\frac1 2} N_1(2\sqrt{\tau}) + A_2\tau^{-\frac1 2} C_1(2\sqrt{\tau}),
\end{equation}
где $A_1, A_2$ - константы. Связь между этими константами находится из краевого условия в точке $z_0$. Отметим, что

\begin{equation}\label{tau}
\sqrt\tau=\sqrt{\frac{S(z)}{h^2}}=\frac1 {2h} \int^z_{z_0} \sqrt{2E(f'^2(z)+1)} dz.
\end{equation}
Подставим (\ref{tau}) в (\ref{psi1}) и получим асимптотическое решение в окрестности точки $z_0$ в виде
$$
u_1(z)=A_1\tau^{-\frac1 2} N_1(\frac{\sqrt{2E}} h \int^z_{z_0} \sqrt{f'^2(z)+1} dz)
+ A_2\tau^{-\frac1 2} C_1(\frac{\sqrt{2E}} h \int^z_{z_0} \sqrt{f'^2(z)+1} dz).
$$
Пусть $A_2=1$; учитывая асимптотики функций Бесселя и Неймана при $\tau\to 0$ (см., например, \cite{Bagrov}), получим
$$
u_1|_{z\to z_0} = -A_1\frac h {\pi\sqrt{2E}} \frac1 {d(z_0,z)} -1+o(1).
$$
Воспользуемся разложением (\ref{razlozh}) и граничным условием (\ref{uslovie})
\begin{equation}\label{A}
A_1=-\frac{\sqrt{2E}\alpha}{2h^3}.
\end{equation}
Подставим (\ref{A}) в (\ref{psi1}) и выпишем асимптотическое решение уравнения (\ref{equ2}), которое лежит в области определения оператора $H$, т.к. удовлетворяет граничному условию (\ref{uslovie})
\begin{equation}\label{reshenie1}
u_1(z)=-\frac{\sqrt{2E}\alpha}{2h^3}\tau^{-\frac1 2}N_1(2\sqrt{\tau})+\tau^{-\frac1 2}C_1(2\sqrt{\tau}),
\end{equation}
где $2\sqrt\tau=\frac{\sqrt{2E}}{h} \int^z_{z_0} \sqrt{(f'^2(z)+1)} dz$.

Поскольку мы рассматриваем случай $m=0$, соответствующая функция $\psi_1(x)$ совпадает с $u_1$ (т.е. не зависит от углов $\theta,\varphi$).

Докажем теперь, что на интервале, где $\psi_1 \ne 0, \psi_2=0$, выполнено
$$
H(e_1\psi_1+e_2\psi_2)=E(e_1\psi_1+e_2\psi_2)+o(h).
$$
Функция $\psi_1=u_1$ удовлетворяет уравнению $H\psi_1=E\psi_1$ с точностью до двух слагаемых, каждое из которых имеет вид
$$
h^2 w(z)\tau F_\tau,
$$
где  функция $w(z)$ голоморфна в окрестности $z_0$, $\tau=\frac{S(z)}{h^2}$. Оценим эту функцию по норме в $L_2$
$$
||h^2w(z)\tau F_\tau||^2_{L_2}\le Ch^4\int^{z_0+\delta}_{z_0} (\tau F_{\tau})^2 w^2 dz \le h^{3-2\epsilon} C_1\delta,
$$
поскольку $|\tau F_{\tau}|\le\frac{Ch^{-\epsilon}}{\sqrt{h}}$. Отсюда
$$
||h^2w\tau F_\tau||\le h^{\frac3 2-\epsilon}\sqrt{C_1\delta}=o(h),
$$
что и требовалось доказать.

Сформулируем результат этого пункта
\begin{lemm}
Функция $\psi_1$ (см. \eqref{reshenie1}) удовлетворяет уравнению $H\psi_1=E\psi_1+o(h)$ в некоторой не зависящей от $h$ окрестности точки $x_0$
\end{lemm}

\subsection{Решение вне окрестности точки $z_0$.}
Уравнение вида (\ref{equ2}) вне окрестности особой точки $z_0$ имеет решение, аналитическое в окрестности точки $z_1$, асимптотика которого с помощью комплексного метода ВКБ строится с точностью до $O(h^2)$ (см. \cite{Fedoruk}). При использовании этого метода условие $H\psi_2=E\psi_2+o(h)$ выполняется по построению. Решение вне сколь угодно малой не зависящей от $h$ окрестности точки $z_1$ представляется в виде (\cite{Fedoruk}):
\begin{equation}\label{psi2}
\psi_2(z)=\nu_1\omega_1(z) + \nu_2\omega_2(z),
\end{equation}
где $\nu_1, \nu_2$ константы, а $\omega_1, \omega_2$ --- фундаментальные асимптотические решения уравнения (\ref{equ2}) при $h\to 0$. А именно
$$
\omega_{1,2}=\omega^o_{1,2}(z)(1+O(h)),
$$
$$
\omega^o_{1,2}(z)=(f'^2(z)+1)^{\frac1 4} \left( \frac{f(z)}{f(z_1)} (\frac{f'^2(z)+1}{f'^2(z_1)+1})^{-\frac1 2} \right)^{-\frac1 2} \exp(\pm i\frac{\sqrt{2E}}{h} \int^z_{z_1} \sqrt{f'^2(z)+1} \, dz).
$$
Найдем константы $\nu_1, \nu_2$ --- это компоненты собственного вектора матрицы монодромии уравнения (\ref{equ2}), соответствующей точке $z_1$. Асимптотика этой матрицы также вычислена в \cite{Fedoruk}; она имеет вид
$$
\begin{pmatrix}
2 & D \\
-(D)^{-1} & 0
\end{pmatrix},
$$

где $D=-\exp\left( \pm \frac{\sqrt{2E}}{h} \int_\beta \sqrt{f'^2(t)+1} \, dt \right)$. Здесь $\beta$ --- замкнутый контур, охватывающий точку $z_1$. Отсюда получаем
$$
\begin{pmatrix}
\nu_1 \\ \nu_2
\end{pmatrix} =
\begin{pmatrix}
-CD \\ C
\end{pmatrix},
$$
где $C$ --- константа. Далее подставим выражения для функций и констант в (\ref{psi2}) и получим асимптотическое решение вне окрестности точки $z_0$
\begin{equation}\label{reshenie2}
\psi_2(z)=\chi(z)\cos\left( \frac{\sqrt{2E}}{h} \int\limits_\beta \sqrt{f'^2(t)+1} \, dt +
\frac{\sqrt{2E}}{h} \int\limits_{z_0}^z \sqrt{f'^2(t)+1 \, dt}\right),
\end{equation}
$$
\mbox{где }
\chi(z)=2C(f'^2(z)+1)^{\frac1 4} \left( \frac{f(z)}{f(z_1)} (\frac{f'^2(z)+1}{f'^2(z_0)+1})^{-\frac1 2} \right)^{-\frac1 2}.
$$

Сформулируем результат этого пункта.
\begin{lemm}
Существует функция $\psi_2(x)$, определенная вне любой не зависящей от $h$ окрестности точки $x_0$ и удовлетворяющая вне такой окрестности равенству
$$
||H\psi_2-E\psi_2||_{L_2}\le A_1 Ch^2,
$$
где $A_1$ --- константа. Вне любой не зависящей от $h$ окрестности точки $z_1$ эта функция имеет вид \eqref{reshenie2}
\end{lemm}

\subsection{Склейка локальных асимптотик. Условие квантования.}

Известны асимптотики функций Бесселя и Неймана при $z\to\infty$ (см., например \cite{Bagrov}):
$$
C_{\alpha}(x) \to \sqrt{\frac2 {\pi x}} \cos(x-\frac{\alpha\pi}2-\frac{\pi}4),
$$
$$
N_{\alpha}(x) \to \sqrt{\frac2 {\pi x}} \sin(x-\frac{\alpha\pi}2-\frac{\pi}4),
$$
подставим их в решение (\ref{reshenie1}), полученное в окрестности точки $z_0$
$$
\psi_1(z)|_{\tau\to\infty}=-\frac{\sqrt{2E}\alpha}{2h^3}[\tau^{-\frac1 2}\sqrt{\frac2{2\pi\sqrt{\tau}}} \sin(2\sqrt{\tau}-\frac{3\pi}4) +\tau^{-\frac1 2} \sqrt{\frac2{2\pi\sqrt{\tau}}} \cos(2\sqrt{\tau}-\frac{3\pi}4)]
+O(\tau^{-\frac3 4}),
$$
преобразуем
\begin{equation}\label{reshenie3}
\psi_1(z)|_{\tau\to\infty} = \sqrt{A^2+1}\,[\frac{\tau^{-\frac3 4}}{\sqrt{\pi}} \cos\left( \frac{\sqrt{2E}}h \int\limits^z_{z_0}\sqrt{f'^2(z)+1}\,dz - \frac{3\pi}4 -arcctgA \right) + O(\tau^{-\frac3 4})],
\end{equation}
где $A=-\frac{\sqrt{2E}\alpha}{2h^3}$.

Приравняв аргументы функций (\ref{reshenie2}) и (\ref{reshenie3}) вне окрестностей точек $z_0$, $z_1$, получим уравнение на спектр, описанное в теореме 1
$$
\tan(\frac1 {2h} \oint\limits_{\gamma}(p,dx)+O(h)) = \frac{2h^3}{\sqrt{2E}\alpha},
$$
где  $(p,dx)=\sqrt{2E(f'^2(z)+1)}\,dz\, ,\gamma$ --- указанный выше цикл на лагранжевом многообразии $\Lambda$. При выполнении этого условия функции $\psi_1$ и $\psi_2$ вне сколь угодно малых окрестностей точек $z_0,z_1$ совпадают $\mathrm{mod} O(h^{3/2-\epsilon})$.

Пусть $(H-E)\psi(x)=r$, оценим остаток $r$ на всем отрезке $[z_0,z_1]$ по норме в $L_2$. Подставим $\psi(x)=e_1\psi_1(x)+(1-e_1)\psi_2$ в уравнение (\ref{main2}).
Имеем
$$
||r||^2 = \int_{z_1}^{z_0}|r|^2 dz \le I_1^2+I_2^2+I_3^2,
$$
где $$
I_1=\int_{z_0}^{z_0+\delta}|r|^2 dz, \,\,
I_2=\int_{z_0+\delta}^{z_1-\delta}|r|^2 dz, \,\,
I_3=\int_{z_1-\delta}^{z_1}|r|^2 dz.
$$

Ранее получено, что $I_1 \le h^{3/2-\epsilon}A_2, I_3 \le h^{2-\epsilon} A_1$, где $A_1, A_2$ - константы. На интервале $(z_0+\delta, z_1-\delta)$ имеем $\psi_1(z)\ne 0, \psi_2(z)\ne 0$ и $\psi_1=\psi_2+O(h^{3/2-\epsilon})$, а значит, по крайней мере $I_2\le h^{3/2-\epsilon}A_3$, где $A_3$ - константа. То есть
$$
||r|| \le \sqrt{h^{3-2\epsilon} A_2^2 + h^{3-2\epsilon} A_3^2 + h^{4-2\epsilon}4 A_1^2} \le h^{3/2-\epsilon}\sqrt{A_2^2 + A_3^2 + h A_1^2} = o(h).
$$
Таким образом, построено решение, такое что
\begin{equation}\label{END}
(H-E)\psi(x)=o(h).
\end{equation}

Для завершения доказательства теоремы 1 воспользуемся следующей леммой
\begin{lemm}
(см., например, \cite{Maslov}) Пусть оператор $H$ - самосопряжен и
$$
(H-E)\psi=f,
$$
где $\psi\in D(H)$, $||\psi||=1$. Тогда существует точка $E_0$, принадлежащая спектру оператора $H$, для которой $|E-E_0|\le ||f||$.
\end{lemm}
Из этой леммы и равенства (\ref{END}) вытекает утверждение теоремы.

\end{document}